# Quadrupolar Glass State in para-hydrogen and ortho-deuterium under pressure.

T.I.Schelkacheva.

*Institute for High Pressure Physics, Russian Academy of Sciences, Troitsk 142092, Moscow Region, Russia*

*e-mail: schelk@hppi.troitsk.ru*

The main features of the possible quadrupolar glass state in ortho-deuterium and para-hydrogen under high pressure are predicted and considered in replica-symmetric approximation in analogy with glassy behavior of diluted ortho-hydrogen at low pressures. The quadrupolar model with $J = 2$ is suggested. The orientational order and glass regime grow continuously on cooling just as it is in the case of ortho-para hydrogen mixtures at zero pressure.

The structure of high pressure phases of solid hydrogen has been widely investigated recently (e.g. [1, 2, 3]). At low pressures, the centers of para-hydrogen and ortho-deuterium molecules occupy the lattice sites of the hcp structure. The molecules are in spherically symmetric states (phase I or SP). Between 28 Gpa for o-$D_2$ and 110 Gpa for p-$H_2$ and ~150 Gpa the molecules become orientationally ordered in broken-symmetry phase II (BS).

High pressure breaks the rotational symmetry of the $J=0$ solids. The ordering occurs even in the systems of para-hydrogen and ortho-deuterium molecules because of mixing of higher order $J$ levels with the ground state $J = 0$ as a result of the increase of the intermolecular interactions at high pressures [1,2]. The anisotropic interaction between two molecules is dominated at low pressures [5] by electrostatic quadrupole-quadrupole (EQQ) interaction, which plays a major role below 100 Gpa [1, 4]. However it is necessary to take into account other anisotropic interactions to understand the results of precise experiments (e.g. Raman scattering [1]).

The purpose of this article is to show the possibility of orientational glass state in solid p-$H_2$ and o-$D_2$ under pressure. Only the main features of this phenomenon will be taken into account. The $J = 0 \rightarrow J = 2$ transitions take place as a result of the pressure increasing. In this case the $J=2$ molecules may be randomly distributed on a close packed lattice. It has been indicated in [2, 6] that rotational disorder under pressure may be frozen. The molecules with $J=0$ are spherically symmetric, have no electric quadrupole moment and play the role of dilutant.

So we can consider the quadrupolar glass state in analogy with glassy behavior of diluted o-$H_2$ and p-$D_2$ at low temperatures and pressures. In this case only the ortho-hydrogen and para-deuterium molecular species are orientable. They have orbital angular momentum $J = 1$. The ordered state is characteriezed by a long-range orientational order at high ortho-$H_2$ and para-$D_2$ concentrations [5]. However, for concentrations less than approximately 55%, measurements show no evidence of orientational phase transition. Instead, NMR experiments have been interpreted in terms of freezing of the the orientational degrees of freedom [7, 5]. This state is so called quadrupolar glass.

The EQQ interaction can bring about orientational glass state at high pressure in p-$H_2$ and o-$D_2$. The rough estimation of $0 \rightarrow 2$ transition probability $\alpha$ can be done [4, 6] using quantum mechanical perturbation theory because the main anisotropic part ( EQQ ) of the intermolecular interaction is small [5]. We have $\alpha = 4 \times 10^{-4}$ for $D_2$ and $\alpha = 6 \times 10^{-5}$ for $H_2$ at zero pressure. The probability to find a $J = 2$ molecule increases strongly with pressure; $\alpha = 0.1$ at 40 Gpa for $D_2$ and 150 Gpa for $H_2$ and $\alpha = 0.4$ for

D$_2$ at 150 Gpa. It is possible that an intermediate range of α exists where a quadrupolar glass occurs.

Let us consider the system of particles on lattice sites $i, j$ with the truncated EQQ Hamiltonian

$$H = -\frac{1}{2}\sum_{i \neq j} J_{ij} Q_i Q_j \tag{1}$$

where the quadrupole component $Q = \frac{1}{2}\left[3\left(\frac{z}{r}\right)^2 - 1\right]$ can be replaced by equivalent operator with the same matrix elements $\hat{Q} \sim [3J_z^2 - J(J+1)]$ in the space $J$ = const. $Sp\hat{Q} = 0$.

Quadrupolar glass freezing in mixtures of ortho- and para-hydrogen has been considered [8] with

$$Q = 3J_z^2 - 2, J_z = 1, 0, -1 \text{ and } J = 1. \tag{2}$$

This model describes well the zero pressure experiments on ortho- para mixtures even in replica-symmetric approach.

Now the Hamiltonian (1) will be examined on the condition $J$ = 2 and $J_z$ = 0, 1, -1, 2, -2.

$$Q = \frac{1}{3}[3J_z^2 - 6], \tag{3}$$

$J_{ij}$ are random exchange interactions with Gaussian probability distribution

$$P(J_{ij}) = \frac{1}{\sqrt{2\pi} J} \exp\left[-\frac{(J_{ij} - J_0)^2}{2 J^2}\right].$$

The scaling $J = \tilde{J}/\sqrt{N}$, $J_0 = \tilde{J}_0/N$ ensures as usual sensible thermodynamic limit. The multiplier 1/3 in (3) is used for simplicity.

Using replica method the free energy is obtained in the form

$$\langle F \rangle_J / NkT = -\lim_{n \to 0} \frac{1}{n} \max \{ -\sum_\alpha \frac{(x^\alpha)^2}{2} - \sum_\alpha \frac{(w^\alpha)^2}{2} - \sum_{\alpha > \beta} \frac{(y^{\alpha,\beta})^2}{2} +$$

$$+ \ln Tr_{\{Q^\alpha\}} \exp\left[ \sum_\alpha x^\alpha \sqrt{\frac{\tilde{J}_0}{kT}} Q^\alpha + \sum_\alpha w^\alpha \frac{1}{\sqrt{2}} t (Q^\alpha)^2 + \sum_{\alpha > \beta} y^{\alpha,\beta} t Q^\alpha Q^\beta \right] \}.$$

Extreme conditions for the free energy give the equations for order parameters

$$m_\alpha = (x^\alpha)^{extr} / \sqrt{\frac{\tilde{J}_0}{kT}} = \langle Q^\alpha \rangle_{eff} ; \quad q^{\alpha,\beta} = (y^{\alpha,\beta})^{extr} / t = \langle Q^\alpha Q^\beta \rangle_{eff} ;$$

$$p^\alpha = (w^\alpha)^{extr} \sqrt{2}/t = \langle (Q^\alpha)^2 \rangle_{eff} ;$$

where $t = \tilde{J}/kT$ and averaging is performed with the effective Hamiltonian

$$-H_{eff} = \sum_\alpha \frac{\tilde{J}_0}{kT} m_\alpha Q^\alpha + \sum_\alpha \frac{t^2}{2} p^\alpha (Q^\alpha)^2 + \sum_{\alpha > \beta} t^2 q^{\alpha,\beta} Q^\alpha Q^\beta$$

Free energy and order parameters in the replica-symmetric [9] case become

$$F = -NkT \left\{ -\left(\frac{\tilde{J}_0}{kT}\right) \frac{m^2}{2} + t^2 \frac{q^2}{4} - t^2 \frac{p^2}{4} + \int_{-\infty}^{\infty} \frac{dz}{\sqrt{2\pi}} \exp\left(-\frac{z^2}{2}\right) \ln Sp[\exp(\theta_2 Q + \theta_1 Q^2)] \right\}. \quad (4)$$

Here $\theta_1 = t^2 \frac{p-q}{2}$ and $\theta_2 = zt\sqrt{q} + m\left(\frac{\tilde{J}_0}{kT}\right)$.

Order parameters are: $m$ – quadrupolar order parameter (analog of magnetic moment in spin glasses), $q$ – glass order parameter and $p$ – auxiliary order parameter.

$$m = \int_{-\infty}^{\infty} \frac{dz}{\sqrt{2\pi}} \exp\left(-\frac{z^2}{2}\right) \frac{Sp[Q \exp(\theta_2 Q + \theta_1 Q^2)]}{Sp[\exp(\theta_2 Q + \theta_1 Q^2)]} =$$

$$= \int_{-\infty}^{\infty} \frac{dz}{\sqrt{2\pi}} \exp\left(-\frac{z^2}{2}\right) 2\left\{\frac{\exp(4\theta_1)[-\exp(-2\theta_2) + \exp(2\theta_2)] - \exp(-\theta_2 \theta_1)}{\exp(4\theta_1)[\exp(-2\theta_2) + 2\exp(2\theta_2)] + 2\exp(-\theta_2 \theta_1)}\right\}, \quad (5)$$

$$q = \int_{-\infty}^{\infty} \frac{dz}{\sqrt{2\pi}} \exp\left(-\frac{z^2}{2}\right) \left\{\frac{Sp[Q \exp(\theta_2 Q + \theta_1 Q^2)]}{Sp[\exp(\theta_2 Q + \theta_1 Q^2)]}\right\}^2 =$$

$$= \int_{-\infty}^{\infty} \frac{dz}{\sqrt{2\pi}} \exp\left(\frac{-z^2}{2}\right) 4\left\{\frac{\exp(4\theta_1)[-\exp(-2\theta_2) + \exp(2\theta_2)] - \exp(-\theta_2 \theta_1)}{\exp(4\theta_1)[\exp(-2\theta_2) + 2\exp(2\theta_2)] + 2\exp(-\theta_2 \theta_1)}\right\}^2, \quad (6)$$

$$p = \int_{-\infty}^{\infty} \frac{dz}{\sqrt{2\pi}} \exp\left(-\frac{z^2}{2}\right) \frac{Sp[Q^2 \exp(\theta_2 Q + \theta_1 Q^2)]}{Sp[\exp(\theta_2 Q + \theta_1 Q^2)]} =$$

$$= \int_{-\infty}^{\infty} \frac{dz}{\sqrt{2\pi}} \exp\left(-\frac{z^2}{2}\right) 2\left\{\frac{2\exp(4\theta_1)[\exp(-2\theta_2) + 2\exp(2\theta_2)] + \exp(-\theta_2 \theta_1)}{\exp(4\theta_1)[\exp(-2\theta_2) + 2\exp(2\theta_2)] + 2\exp(-\theta_2 \theta_1)}\right\}. \quad (7)$$

The temperature dependence of order parameters obtained from (5) – (7) is represented in figures 1 - 3. There is no trivial solution $m = 0$, $q = 0$ at finite temperature because $SpQ^3 \neq 0$. The orientational order and glass regime grow continuously on cooling just as it is in the case of the Hamiltonian (1) – (2) describing the ortho-para hydrogen mixtures at zero pressure [8]. The quadrupolar long range order is present for $T > 0$ even if $J_0 = 0$. In the pure case $(\tilde{J} = 0, \tilde{J}_0 \neq 0)$ we have from (5) – (6) $q^{1/2} = m$.

It is easy to see from (2) that $Q^2 = 2 - Q$. So the equation for $p$ is not independent and $p = 2 - m$. There is no similar expression for $Q^2$ from (3) and for order parameter $p$ here.

Using the equation (4) for the free energy the heat capacity can be written in the form

$$\frac{C_v}{kN} = \frac{d}{d(kT/\tilde{J})}\left\{\left(\frac{\tilde{J}}{kT}\right)\frac{(q^2 - p^2)}{2}\right\} - \left(\frac{\tilde{J}_0}{\tilde{J}}\right)m\frac{dm}{d(kT/\tilde{J})}. \quad (8)$$

Specific heat as a function of ($kT/\tilde{J}$) calculated from (8) for the four cases of ($\tilde{J}_0/\tilde{J}$) is shown in figure 4. The dependence of $C_v$ on the temperature is smooth as in ortho-para mixtures at zero pressure.

In conclusion, the main features of the random quadrupolar system (1), (3) are considered in the replica-symmetric mean-field approximation. This system with $J = 2$ has not been considered earlier. The possible realization of the quadrupolar glass state under high pressure in ortho-deuterium and para-hydrogen is predicted.

The author is grateful to E.E.Tareyeva and V.N.Ryzhov for useful and stimulating discussions. This work was supported by the Russian Foundation for Basic Research (Grant 02-02-16622-a).

Fig.1. Order parameters for the case $\widetilde{J}_0/\widetilde{J} = 0$.

Fig.2. Order parameters for the case $\widetilde{J}_0/\widetilde{J} = 1$.

Fig.3. Order parameters for the case $\widetilde{J}_0/\widetilde{J} = 2.5$.

Fig.4. Specific heat as a function of $(kT/\widetilde{J})$ for the cases $\widetilde{J}_0/\widetilde{J} = 0; \widetilde{J}_0/\widetilde{J} = 1; \widetilde{J}_0/\widetilde{J} = 1.4; \widetilde{J}_0/\widetilde{J} = 2.5$

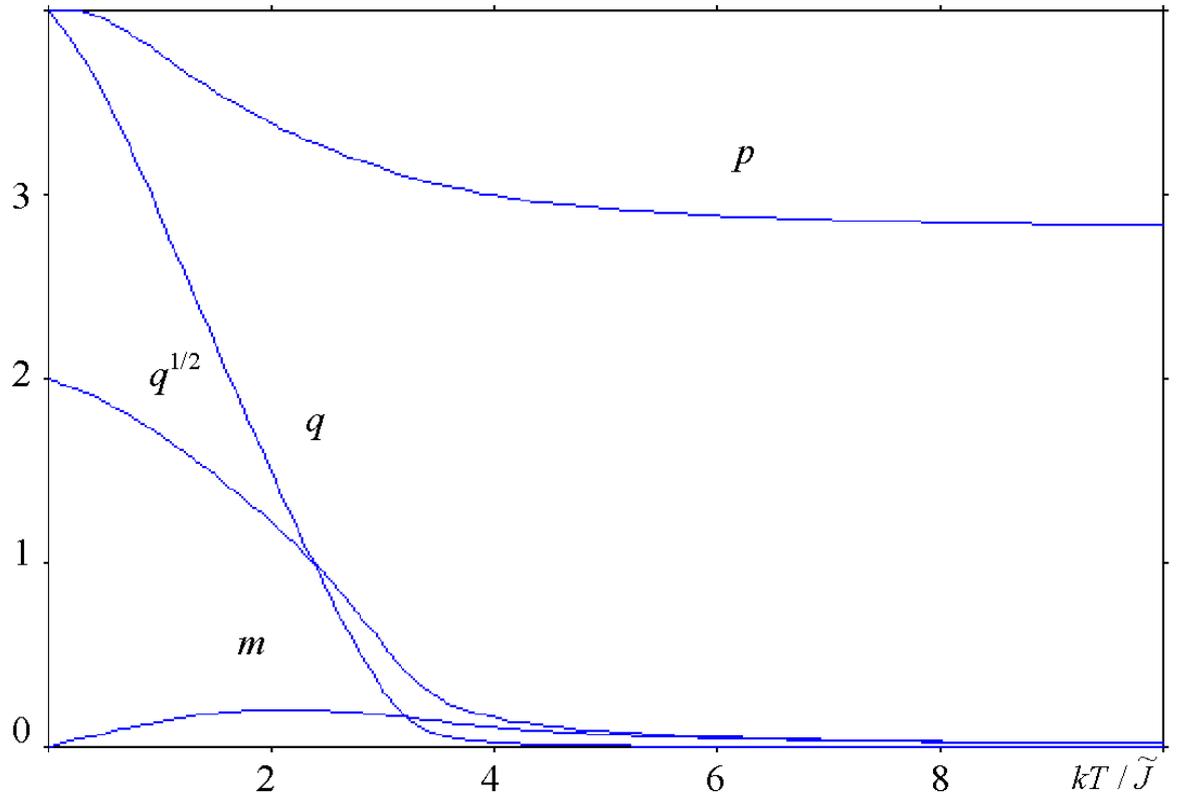

Fig. 1

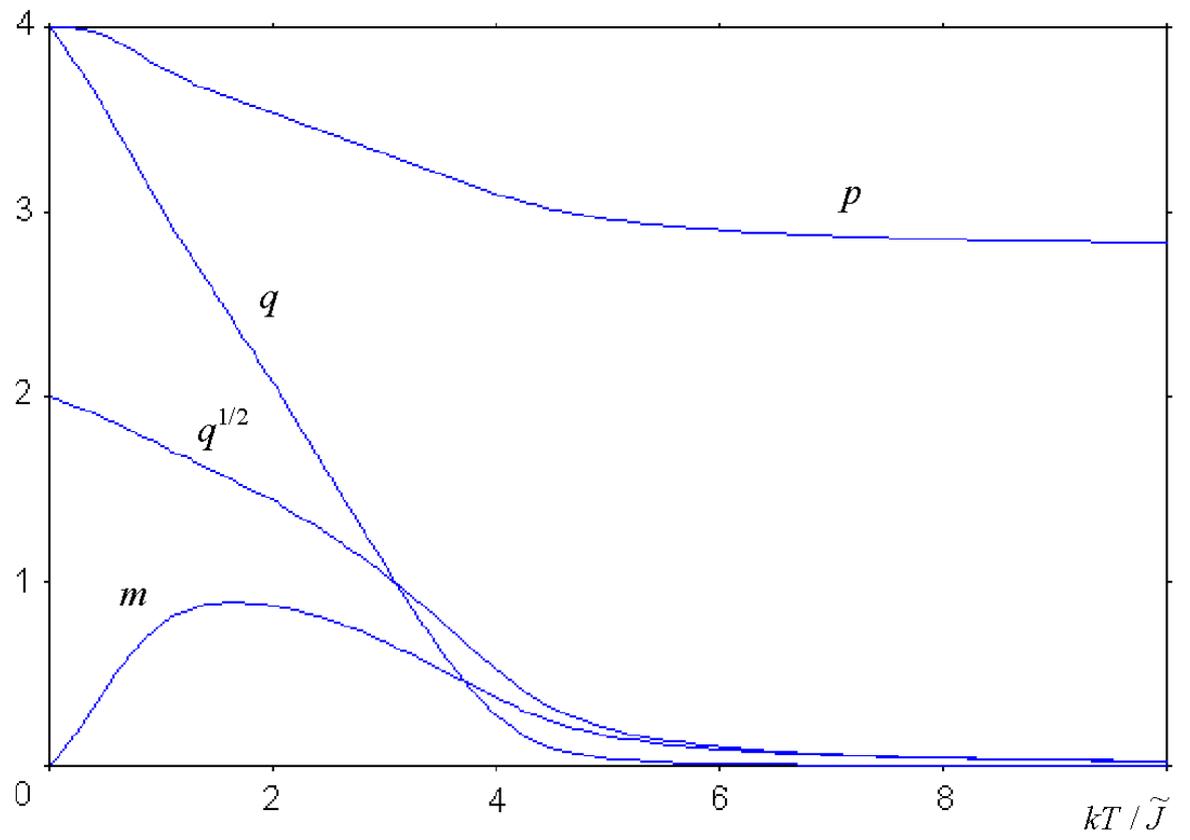

Fig. 2

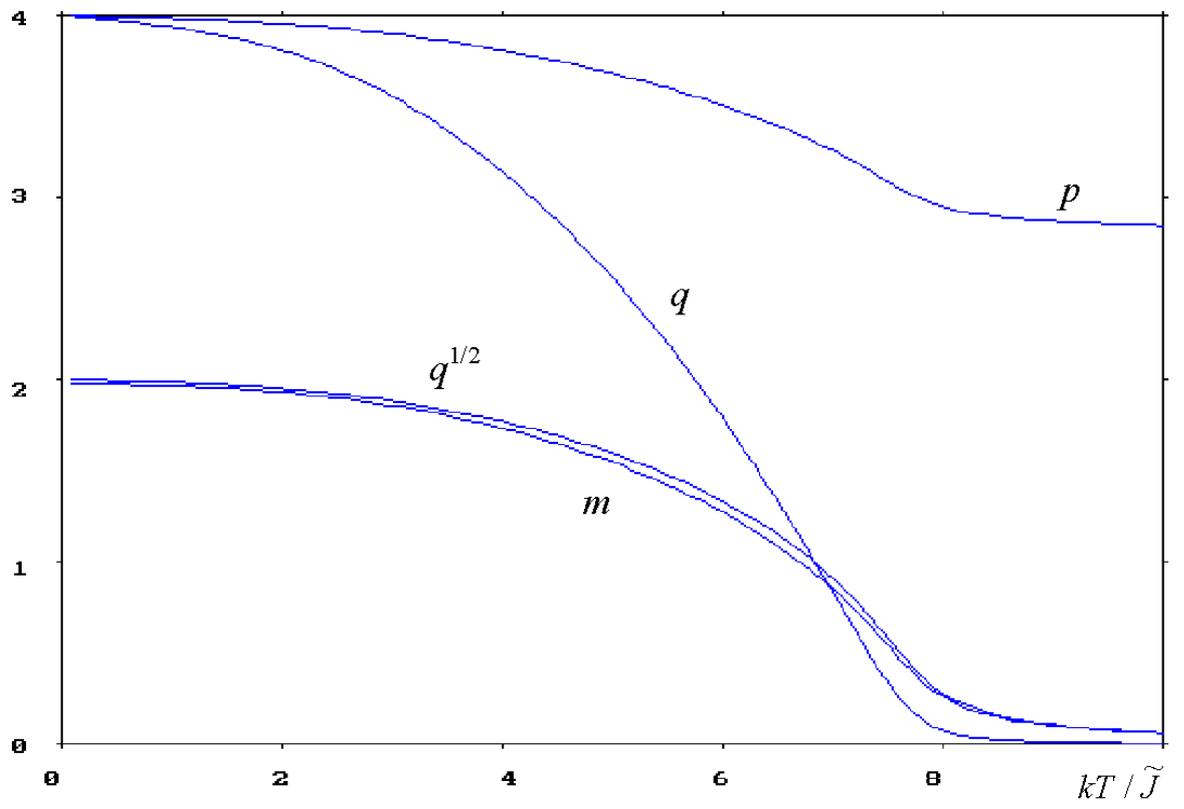

Fig. 3

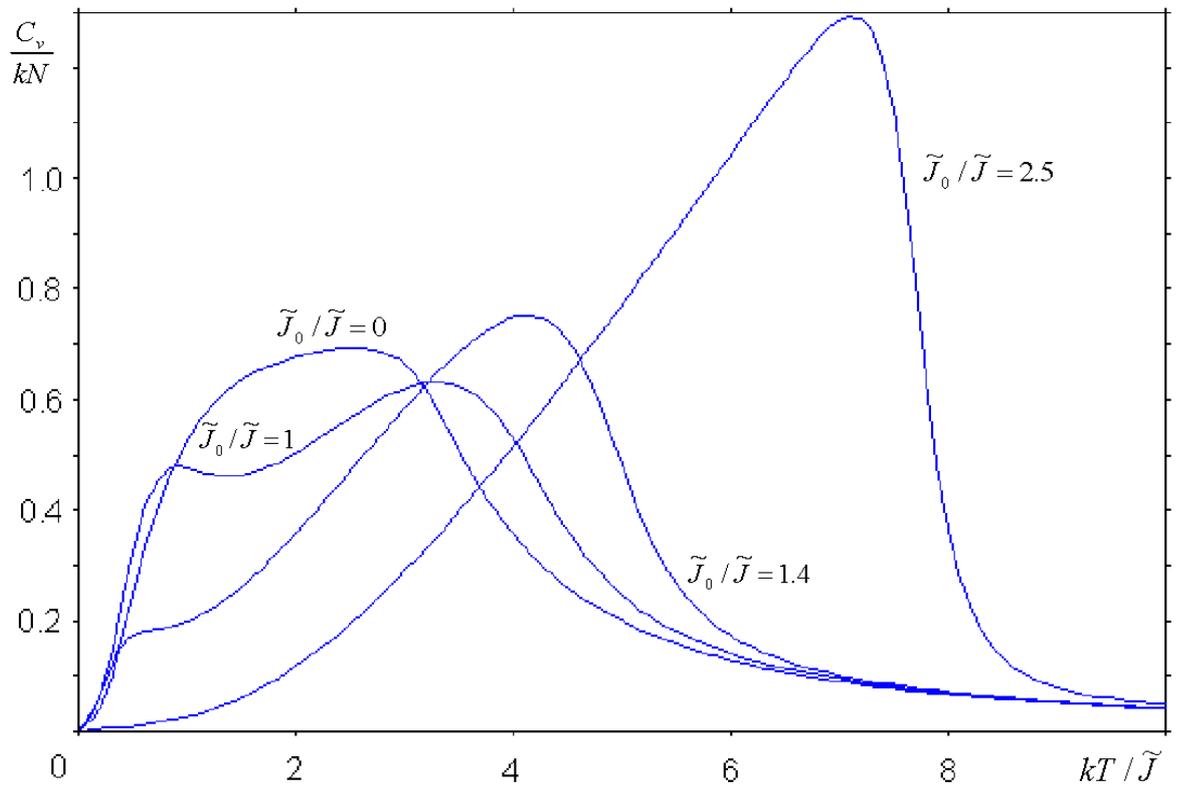

Fig. 4